# *FixMyStreet* Brussels: Socio-Demographic Inequality in Crowdsourced Civic Participation

Burak Pak [a], Alvin Chua [b] and Andrew Vande Moere[b]

[a]Faculty of Architecture, KU Leuven, Leuven, Belgium; [b]Department of Architecture, KU Leuven, Leuven, Belgium

**ABSTRACT**

*FixMyStreet* (*FMS*) is a web-based civic participation platform that allows inhabitants to report environmental defects like potholes and damaged pavements to the government. In this paper, we examine the use of *FMS* in Brussels, the capital city of Belgium. Analyzing a total of 30,041 reports since its inception in 2013, we demonstrate how civic participation on *FMS* varies between the ethnically diverse districts in Brussels. We compare *FMS* use to a range of sociodemographic indicators derived from official city statistics as well as geotagged social media data from Twitter. Our statistical analysis revealed several significant differences between the districts that suggested that crowdsourced civic participation platforms tend to marginalize low-income and ethnically diverse communities. In this respect, our findings provide timely evidence to inform the design of more inclusive *crowdsourced*, civic participation platforms in the future.

## Introduction

Recent advances in crowd-based production technologies and other associated methods have triggered numerous crowdsourcing (Howe, 2006) initiatives for a wide range of government, civic, and commercially oriented causes. In the practice of urban governance, crowdsourcing is typically employed to understand the concerns of citizens and engage them in collaborative decision-making (Seltzer and Mahmoudi, 2013). Web-based platforms that crowdsource civic participation for urban governance such as *FixMyStreet* (FixMyStreet, 2016), *PublicStuff* (PublicStuff, 2016) and *SeeClickFix* (SeeClickFix, 2016) in particular, have become increasingly prominent in recent years (Atzmanstorfer and Blaschke, 2013; crowdgov, 2016). Conceived as public service improvement systems to promote transparency and accountability, these platforms allow citizens to work collaboratively with their governments to sustain their environments. Citizens report defects to their governments through these platforms and governments commit to respond in a timely manner. The reported defects are aggregated on a publicly accessible map that reflects the state of the environment. Using these maps as monitoring tools, citizens can track and discuss the performance of their governments in response to the overall state of their environments. Though crowdsourced civic participation is generally perceived as a positive mechanism for empowering ordinary citizens and democratizing an otherwise bureaucratic process (Martínez et al., 2009; Wallack and Nadhamuni, 2009), soliciting civic participation by way of web-based platforms has drawn substantial criticism for marginalizing certain segments of the population (Brabham, 2008).

Because of the ubiquity of the Internet, it is assumed that web-based civic participation platforms lower the barriers for citizens to engage with their governments (Baykurt, 2012). Yet, we believe that the inclusion (Jaeger et al., 2012) of people from different socio-demographic backgrounds remains a substantial challenge. Access and use of these platforms, in particular, might be difficult for marginalized groups in society (Pak and Verbeke, 2014) as it requires skills and aptitude that are often out of their reach (Haklay, 2013). Furthermore, the cost of the equipment to access the platforms (such as smartphones, tablets, or computers) is a potential factor promoting exclusion. Thus, there is a need to better understand the degree of socio-demographic inequality on these platforms. Our specific research questions are: (RQ1) Are citizens of varying income levels represented on web-based civic participation platforms? (RQ2) Are citizens of different ethnicities represented on web-based civic participation platforms? For this purpose, we analyzed historical usage logs from a popular web-based civic participation platform to investigate how civic participation varies between people of different income levels and ethnicities. To validate our results, we further compared the hypothesized variations in civic participation to traces of Internet access approximated by the use of online social media. This analysis was based on a combination of historical usage logs from a web-based civic participation platform, geotagged social media data, and official city statistics. A set of socio-demographic indicators were derived from these datasets and analyzed to arrive at three hypotheses that are directly related to our research questions.

The paper is structured as follows. We begin with a background review on applications of crowdsourcing in urban governance. This is followed by a description of our methodology, elaborating on how our socio-demographic indicators were calculated, visualized, and statistically compared. We then describe the results of our analysis and discuss the implications of our results in relation to the use of web-based civic participation platforms in general. Finally, we conclude with several suggestions on how the challenges posed by socio-demographic inequality can be addressed in future implementations of web-based civic participation platforms.

## Background

Crowdsourcing is described as the act of outsourcing a traditional task by making an open call to an undefined but large group of people (Howe, 2006). While this definition has evolved over the last decade to cover a wide range of applications, what is common among crowdsourcing initiatives is the way in which participatory actions are organized for mutual benefit. The crowd voluntarily performs a task and receives satisfaction in various forms (i.e. social recognition, skill development, economic reward, etc.). Similarly, the crowdsourcer benefits from what the crowd has produced and then uses the outcome (Brabham, 2008; Estellés-Arolas and González-Ladrón-de-Guevara, 2012).

### Crowdsourcing Civic Participation with Web-Based Platforms

Over the last decade, numerous web-based civic participation platforms have been developed to support various types of urban planning, design, and governance tasks. *311Online* (NYC311, 2016) implemented by New York City, for instance, was among one of the earliest feedback systems for reporting complains and incidents. Likewise, *OpenPlans* (*OpenPlans*, 2016), a web mapping tool has a history of engaging communities in various urban planning projects like the improvement of bicycle network infrastructure and placement of urban tree canopies. Other platforms like *Ushahidi* (*Ushahidi*, 2016) developed by a non-profit software company in Nairobi with the same name and *OpenDataKit* (*OpenDataKit*, 2016) by a group of researchers at the University of Washington's Department of Computer Science also provide citizens with interactive mapping interfaces to aggregate incident reports and organize collective responses. *MySociety* "charity" (*mySociety*, 2016) in particular, has produced numerous platforms for a wide range of civic activism in the United Kingdom. For example, *TheyWorkForYou* allows citizens to monitor their parliamentary representatives (*TheyWorkForYou*, 2016), while *WhatDoTheyKnow* allows citizens to request information from public authorities (*WhatDoTheyKnow*, 2016). Similarly *WriteToThem* enables citizens to formally address parliamentary representatives, councilors, and members of the European Parliament (*WriteToThem*, 2016).

The most promising aspect of crowdsourcing for urban governance lies in the ability to address problems that are typically large-scale, complex, and difficult to resolve in a top-down manner (Surowiecki, 2005). Unlike the expert-driven approach to problem solving, crowdsourced solutions emerge from the collective efforts of multiple motivated amateurs (Brabham, 2008). For example, defects in the urban environment may be costly for the government to systematically track from the top down, yet may be easily catalogued with the collective efforts of citizens from bottom-up. This is especially pertinent for urban governance since it allows for large numbers of people, with deep knowledge of the environment, to be actively involved in improving the quality of their lives (Erickson, 2010). Furthermore, the crowdsourced outcomes are derived from citizens who actually use the space (Brabham, 2009) and are more likely to be relevant for a wider public instead of specific interest groups (Brabham, 2008).

Though web-based platforms hold enormous potential to garner civic participation, they are also likely to face challenges rooted in the digital divide (Pippa, 2001). Two studies in particular shows that these platforms have been biased towards citizens with better education and higher income (Helsper, 2008; Escher, 2011). Presumably, these citizens were already interested in governance issues (Baykurt, 2012) and make use of these platforms to voice their dissatisfaction with "matters of everyday service delivery" (Verplanke et al., 2010: 187). In stark contrast, citizens without access the Internet miss the opportunity to participate. Especially in developing countries and developed countries with high income inequality, the cost of the hardware stands as a potential challenge to accessibility. Digital literacy is another obvious limitation for these platforms. Ethnic minorities may not understand the language of instruction while the visual interface may be daunting for those without the necessary skills and experience to navigate (Baykurt, 2012). Naturally, public perception of these platforms also influences their potential to succeed (Pak and Verbeke, 2014). How governments deal with problems as well as the measures taken to reach certain ends will inevitably determine the amount of trust citizens have in these platforms (Offenhuber, 2015). It is also important to establish and maintain good participation rates, as citizens may not be willing to participate in initiatives that they consider to have failed (Brabham, 2009).

## FixMyStreet *and Relevant Research Findings*

*FixMyStreet* (*FMS*) is a popular web-based civic participation platform that has been adopted in many cities worldwide (*mySociety*, 2016). It is one of the first projects that tested the potential of crowdsourcing. It promotes civic engagement by providing the public with a channel for reporting environmental defects like potholes in the tarmac, broken paving slabs and defective street lighting. Citizens can also use the platform to track specific defects or participate in discussions on governance issues. The main features of *FMS* include statistics and search events by municipality; subscription to alerts of events that appear in a particular place; submission of photos about events; and discussions of reports (Santos et al., 2013).

Until recently, relatively little work has been done on the subject of *FMS* usage behavior. The earliest research on this topic was by King and Brown (2007) who used a limited dataset of 196 reports and 3,982 updates on these reports. This paper is in the form of a case study with interviews of 36 stakeholders. It surveyed a quite early form of the *FMS* platform that was rather different from the *FMS* today; it was missing most of the features listed above (such as discussion and a dynamic map). As a result, the researchers concluded that *FMS* was easy to use but there was a further need for community building. However, despite the claims of *FMS* to enable citizens to report, view, or discuss local problems, there was very little public debate taking place. The researchers noted that *FMS* intended to place information in the hands of citizens but at the same time limited the accuracy and variety of information flows, such as not allowing citizens to discuss what anti-social behavior is.

One of the few in-depth studies that we are aware of (Sjoberg et al., 2015) revealed that a successful first experience on the platform, like reporting a pothole and having it fixed, contributes to a 54 percent increase in the probability of further participation. Likewise, knowledge of participant socio-demographics is sparse. Results of another study conducted in the United Kingdom (Gibson et al., 2014) indicated that participants from various education levels were equally represented on the platform, though a majority of these participants were white. The number was drastically low when compared to official census data. Only 2 percent of the participants on *FMS* were determined to be ethnic minorities whereas the actual figure should have been 13 percent. Accordingly, comparison to other crowdsourcing platforms revealed that *FMS* is specifically prone to demographic bias. Participation of ethnic minorities was gauged to be an average rate of 2 percent on *FMS* while 4.9 percent was recorded on other platforms. The study suggests that the lack of interest on the part of ethnic minorities could have been due to their social attitudes towards involvement with local communities as well as their perception about their ability to influence policies. Another potential cause is the relatively complicated interface of *FMS*, which requires mapping skills.

While these studies revealed the complexity involved in crowdsourcing civic participation with *FMS*, the current understanding of socio-demographic inequality on the platform as well as its possible effects are still limited.

### Brussels Official Statistics Dataset (BISA)

The Brussels official statistics dataset (BISA, 2016; Wijkmonitoring, 2016) is collected by the Brussels Capital Region and includes measures for ethnicity, average age of the population, and the annual taxable income per capita, among other socio-demographic indicators. While we consider this dataset to be a reliable source of information (Kitchin, 2014a) to contextualize participant behavior on *FMS* (Crampton et al., 2013), we also recognize that it is limited in several ways. First, the data are aggregated at the district level and do not provide a high degree of geographic precision (Wijkmonitoring, 2016). Next, the data are compiled on the basis of officially registered citizens, but commuters are unaccounted for. Finally, the indicator for ethnicity is binary and does not take multi-cultural identities into consideration. To address these discrepancies, we derived an alternative indicator of ethnicity based on Geotagged Twitter Data.

### Geotagged Social Media Data as an Alternative Source of Information

The use of social media services has grown tremendously over the last decade, and the impact of such growth has expanded in unexpected ways. Among such services, Foursquare, Instagram, and Twitter are large-scale data sources that may be used to study human behavior in cities (Kitchin, 2014b). Twitter data, in particular, can be considered a repository of social information because the information it captures reflects real-life events in many ways (Java et al., 2007). As large volumes of geotagged Tweets become increasingly available, new opportunities arise to study these data for insights into the spatial behavior of people. Previous works have attempted to study daily urban routines (Naaman et al., 2012), to understand travel patterns at different scales (Chua et al., 2014), to track land patterns use over time (Frias-Martinez et al. 2012), to characterize urban locations (Kling and Pozdnoukhov, 2012; Cranshaw et al., 2012), and to trace lifestyle patterns on the basis of activities (Andrienko et al., 2013). Visualizations depicting the geographic distribution of Tweet languages (Shane, 2014; spatial.ly, 2014) is of particular relevance to our work since this aspect of geotagged Tweets provides the opportunity to sense the whereabouts of ethnic communities by employing language as a proxy.

Twitter is a useful means of acquiring timely and detailed geographic information at very low cost, yet it is biased to assume that the data will naturally yield some value to urban governance without a clear understanding of its limitations. User penetration is a key factor that must be taken into consideration as individuals who post on Twitter represent a privileged demographic with access to the Internet. Of this demographic, studies have shown that Twitter users significantly over-represent densely populated, urban regions (Mislove et al., 2011; Hecht and Stephens, 2014), are predominantly male, and are ethnically skewed (Mislove et al., 2011). Sample validity is another important consideration as Twitter lacks documentation concerning what and how much data are provided through their API. Twitter allows the retrieval of up to 1 percent of all the data, and studies have shown that geotagged Tweets account for only 3.17 percent of that sample. However, this number is shown to be highly representative of the entire corpus (Morstatter et al., 2013) and a growing body of evidence has suggested that geotagging is becoming more of a mainstream activity (Hecht and Stephens, 2014). Reliability is also an important factor to consider when analyzing Tweets (Nabian et al., 2013). Previous studies have shown that 34 percent of Twitter users did not provide real information about their location of origin. Instead, fake locations or sarcastic comments were often incorporated into the data to fool commonly used data analysis tools (Hecht et al., 2011). Other instances of unreliable data include inaccurate geographic and temporal information resulting from weak connection to the GPS network, incorrect settings on an individual's phone, spam-like generic Foursquare check-ins, or automatically generated nike+ logs and missing data that may have been intentionally omitted by the user. Moreover, age is another important challenge for considering the representativeness of Twitter. It is well-documented that Twitter is more popular among younger people, especially among the 18–29 and 30–49 age groups (Duggan, 2015).

## Context

Brussels is a diverse city with a population of approximately one million. Of this number, 67.9 percent are of foreign origin (Hertogen, 2015). It is also a socially segregated city where marginalized ethnic minorities are confined to deprived, low-income districts (BISA, 2016; Wijkmonitoring, 2016). Sint-Joost-ten-Node for instance, is one of the most segregated municipalities in Brussels. Encompassing four densely populated districts, 96 percent of its inhabitants are of foreign origin and 33 percent of this number are not Belgian citizens. Accordingly, official statistics show that the average income per capita of €8,242 in this municipality is less than half of the €16,650 Belgian average (See Figure 1(a)) (STATBEL, 2014; BISA, 2016; Wijkmonitoring, 2016). Similar patterns have been observed throughout the northwestern sector of the city where the majority of the inhabitants are of North African, Turkish, and Sub-Saharan ethnicities (Kesteloot and Meys, 2008). Locals colloquially refer to this region of Brussels as the "disadvantaged croissant" since it takes the shape of a crescent (See Figure 1(b) and 1(c)).

In order to address the emerging complexities of governing the city, local authorities introduced a smart city initiative in 2013 to "encourage its inhabitants, businesses, and administrations to devise new ways of sharing data and projects to improve the efficiency of services and focus them on the needs of users" (CIRB, 2016: online). Accordingly, *FMS* was made available to citizens and local administrative units as the initial step in this strategy (See Figure 2). Since its inauguration, *FMS* Brussels has been portrayed numerous times in the regional and local media as a platform for empowerment (Colleau, 2013; nieuwsblad.be, 2014; Vandenbergh, 2016). The communication channels used for promotion heavily relied on print media, newspapers, and online news sites in French and Dutch which are not necessarily accessible to people of all socioeconomic levels and national origins.

As has already been noted, previous research strongly suggests that certain socio-demographic groups might be marginalized by the use of web-based civic participation platforms. Therefore, we carried out an in-depth analysis of historical usage logs derived from *FMS* Brussels to study the degree of socio-demographic inequality on these platforms. In line with the research questions, background review, and summary statistics from exploratory analysis described later in the paper, our hypotheses are:

- **Research Question RQ1:** Are citizens of varying income levels represented on web-based civic participation platforms?

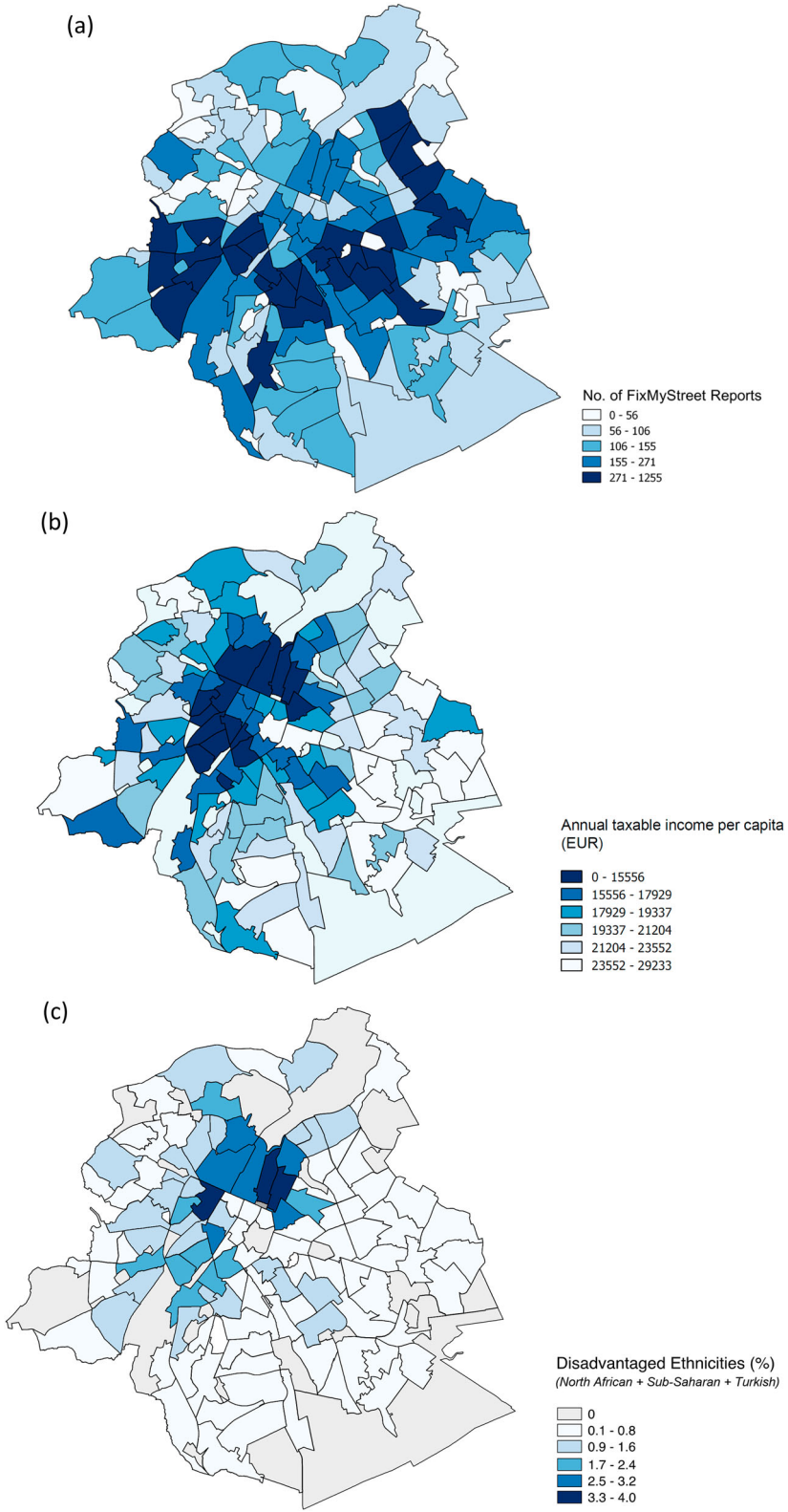

**Figure 1.** Geographic Distribution of (a) The total number of *FixMyStreet* reports (mid-February 2013 to mid-June 2015) (b) The annual taxable income per capita (c) North African, Sub-Saharan and Turkish ethnicities at district levels in the Brussels Capital Region (BISA, 2016; *FixMyStreet*, 2015)

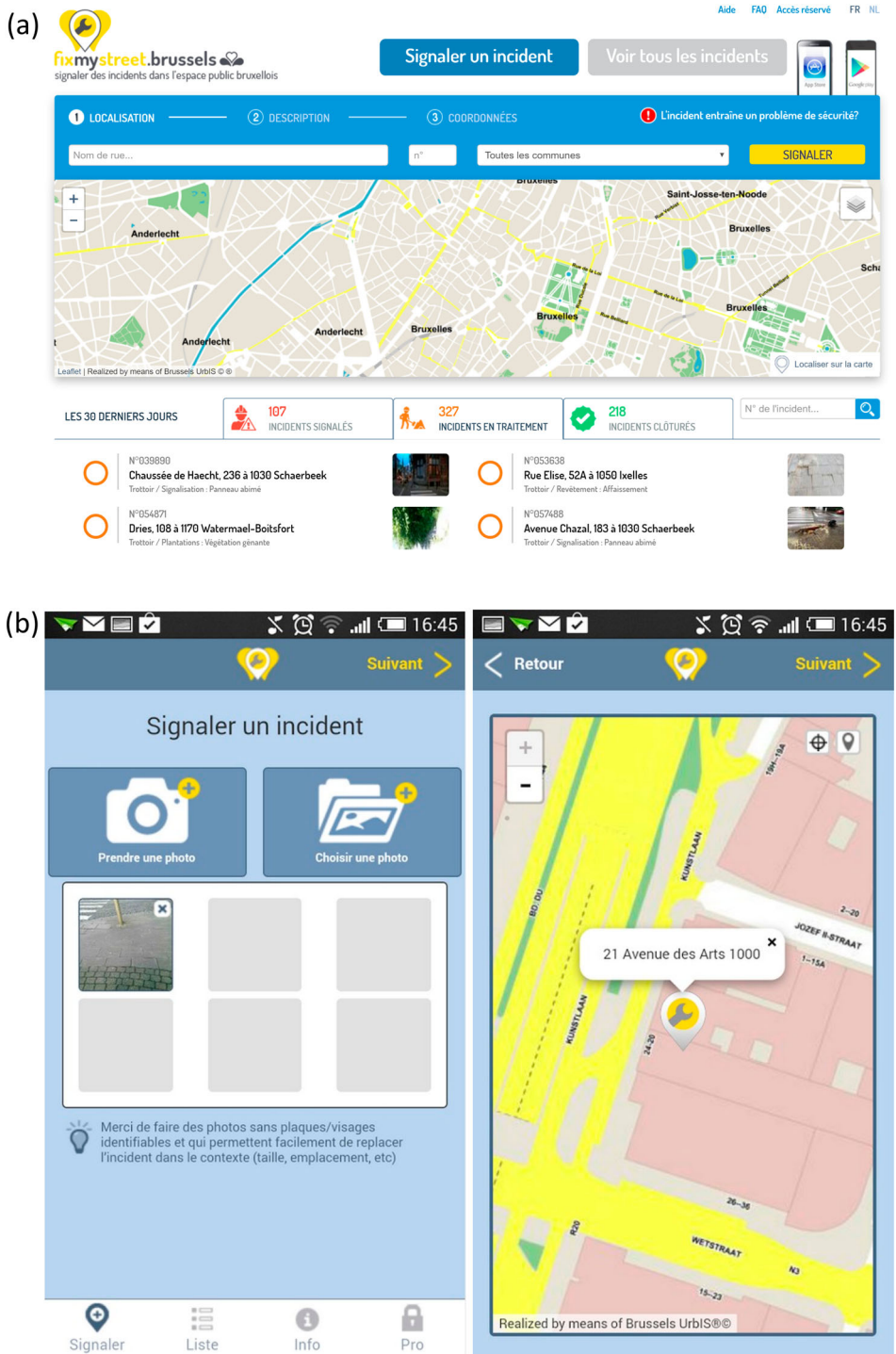

**Figure 2.** Screenshots from the French version of the *FixMyStreet* Brussels Platform (a) web interface (b) Android Smartphone App.

- **Hypothesis:** Citizens from low- and high-income districts should be equally represented on FMS
- **Null H1:** There is no significant difference between average income and rate of *FMS* participation per district.
- **Research Question RQ2:** Are citizens of different ethnicities represented on web-based civic participation platforms?
- **Hypothesis:** Citizens from more or less ethnically diverse districts should be equally represented on FMS.
- **Null H2:** There is no significant difference between ethnic composition (percentage of ethnicity) and the rate of *FMS* participation per district.

## Method

This study employed a combination of exploratory and confirmation analysis on three distinct datasets to test the hypotheses described above:

(1) Historical *FMS usage logs* spanning 29 months from mid-February 2013 to mid-June 2015. The dataset was publicly accessible (*FixMyStreet*, 2015) and downloaded on June 16, 2015. It comprises 30,041 reports posted by 3,511 users. Each report consists of 13 variables among which, five are central for this study: The geographic point coordinate of the reported fault, the unique identifier of the user who reported the fault, when the report was made, when the government responded to the report, and when the fault was resolved.

(2) *Official Statistics* are provided by the city at the district level. The data are organized into 145 districts designated by a Belgian inter-university consortium (BISA, 2016; Wijkmonitoring, 2016). It covers 118 residential areas where 99.7 percent of the population reside, six industrial areas, 18 green areas, and three cemeteries. We used statistical indicators for the share of non-Belgian nationalities (grouped per country), population density, and the average income per capita in 2012 per district.

(3) A set of geotagged Tweets was posted between January 1, 2012 and December 31, 2013, geographically located within the administrative boundary of Brussels. This dataset was obtained from the twitter stream API and consists of 1,134,111 Tweets posted by 36,682 users. The Tweets are categorized into languages with a purpose-built classification procedure. The resulting classification serves as an alternative indicator for the geographic distribution of ethnic groups in Brussels.

The datasets are formatted in different geographic scales and data processing is required for them to be comparable. To this end, we aggregated FMS usage logs and Geotagged Tweets from geographic point coordinates to district-level geographic polygons that are consistent with the Official Statistics. This was followed by exploratory analysis for a general overview of the patterns in datasets. Specifically, we charted the growth of *FMS* Brussels over its lifespan and compared geographic variations in how the platform was used across the city. This preliminary step was paramount in formulating the hypotheses described above and identifying the following biases in the data:

- *Population Density.* Districts with larger populations have the probability of attracting more users who will inevitably generate more reports.
- *Land Area.* Districts with large land areas tend to encompass a greater number of streets and are likely to encounter more problems.
- *Highly Active Users.* Enthusiastic users may submit more reports than an average user. The reports may also be dispersed over a large area.

Normalization is required to overcome these biases and demonstrate a fuller expression of how *FMS* use varies between different districts in Brussels. For this purpose, we adopted 12 normalized indicators of which 11 were computed. Table 1 provides a description of these indicators, the dataset from which they were derived, and the corresponding equations required to compute them. No computation was required for $I$, the average income per district. In the following sections, the value $n$ refers to the number of districts in the city.

### Indicators for FixMyStreet *Usage Data*

The normalized indicator $R_{u.km}$ denotes the rate of civic participation on *FMS* in a given district. This indicator is computed with Equation (1). The value $R$ denotes the original number of reports as provided in the data, while $U$ denotes the total number of users and $Ln$ the total street length per district. Districts with large $R_{u.km}$ values indicate a high rate of civic participation on *FMS*.

$$R_{u.km} = \frac{R}{(U/Ln)} \tag{1}$$

### Indicators for Ethnicity

The normalized indicators for ethnicity are derived from *Official Statistics*. The value $D$ in Equation (2) denotes the aggregate population of disadvantaged ethnicities per district.

**Table 1.** Summary of the 12 standardized indicators computed for each district in Brussels.

| Indicator | Description | Dataset | Equation |
|---|---|---|---|
| $R_{u.km}$ | Number of reports relative to the total number of users and total street length. | FixMyStreet Usage Logs | (1) |
| $D$ | Percentage of population who are from North African, Turkish, or Sub-Saharan ethnicities. | Official Statistics | (2) |
| $M'$ | Percentage of North African ethnicity relative to the total North African population. | Official Statistics | (3) |
| $T'$ | Percentage of Turkish ethnicity relative to the total Turkish population. | Official Statistics | (3) |
| $S'$ | Percentage of Sub-Saharan ethnicity relative to the total Sub-Saharan population. | Official Statistics | (3) |
| $I$ | Average income | Official Statistics | N/A |
| $Ar_{t.u}$ | Percentage of Arabic Tweets relative to the total number of Twitter users who tweet in Arabic. | Geotagged Tweets | (4) |
| $Tr_{t.u}$ | Percentage of Turkish Tweets relative to the total number of Twitter users who tweet in Turkish. | Geotagged Tweets | (4) |
| $SoSw_{t.u}$ | Percentage of Somali and Swahili Tweets relative to the total number of Twitter users who tweet in either language. | Geotagged Tweets | (4) |
| $AD_{t.u}$ | Percentage of aggregated Arabic, Somali and Swahili tweets relative to the total number of Twitter users that tweet in any of the aforementioned languages. | Geotagged Tweets | (5) |

Accordingly, the values $M$, $S$, and $T$ refer to the population of North African, Sub-Saharan, and Turkish citizens respectively. Districts with large $D$ values indicate high concentrations of inhabitants from disadvantaged ethnic backgrounds.

$$D = \frac{(M + S + T)}{\sum_{i=0}^{n} (M + S + T)} \qquad (2)$$

Separate normalized indicators for the distribution of North African $M'$, Sub-Saharan $S'$, and Turkish $T'$ ethnicities are computed with Equation (3). The value $E'$ in Equation (3) corresponds to the indicator for a specific ethnicity while $E$ refers to the total population of that ethnicity in each district. Districts with high $M'$, $S'$, or $T'$ values indicate a high population of North African, Sub-Saharan, and Turkish inhabitants respectively.

$$E' = \frac{E}{\sum_{i=0}^{n} P} \qquad (3)$$

### Alternative Indicators for Ethnicity

We developed an alternative indicator for ethnicity based on the distribution of language use in the city. In this instance, Tweets written in a certain language serve as a proxy for the ethnic population in a given district. This was achieved by way of a three-step classification process. First, empty or automated Tweets were filtered from the dataset. The remaining tweets were then broadly classified on the basis of character encoding (Shuyo, 2016a) and finally refined for precision with a short-text language detection model (Shuyo, 2016b). The classification process yields a probability $p$ for the likeliness of a Tweet to be written in a certain language. Tweets with $p < .75$ were excluded from further data processing to ensure that the classification had a higher degree of precision. The normalized indicator for a certain language was then computed with Equation (4), where $L_{t.u}$ denotes the percentage of Tweets written in that language $L_t$ relative to the total number of Twitter users $L_u$ who posted those Tweets. Districts with high $L_{t.u}$ values indicate a greater concentration of Tweets written in a specific language. Correspondingly, this suggests the presence of citizens who use that language.

$$L_{t.u} = \frac{(L_t/L_u)}{\sum_{i=0}^{n} (L_t/L_u)} \qquad (4)$$

An aggregate indicator for disadvantaged ethnicities $AD_{t.u}$ is computed in a similar manner. This is shown in Equation (5), where the values $Ar_t$, $SoSw_t$ and $Tr_t$ refer to the total number of Tweets written in Arabic, Somali, and Swahili as well as Turkish, while the values $Ar_u$, $SoSw_u$ and $Tr_u$ denote the total number of users who wrote those Tweets. Districts with high $AD_{t.u}$ values indicate high concentrations of Tweets written in the language of that specific ethnicity. Accordingly, this suggests a strong presence of citizens from that ethnicity.

$$AD_{t.u} = \frac{(Ar_t + SoSw_t + Tr_t/Ar_u + SoSw_u + Tr_u)}{\sum_{i=0}^{n} (Ar_t + SoSw_t + Tr_t/Ar_u + SoSw_u + Tr_u)} \qquad (5)$$

## Results

We commence with a description of the geographic and temporal patterns obtained from exploratory analysis of FMS usage logs. We then examine the Official Statistics on ethnicity and income as well as the distribution of ethnic languages captured by Geotagged Tweets. Finally, we elaborate on the results of our statistical comparison based on the hypotheses described in the former section.

### Exploratory Analysis of FMS Usage Logs

*Temporal analysis.* Figure 3 depicts a time series visualization of FMS usage logs over its lifespan of 29 months. It illustrates variations in monthly activity, comparing the number of new and existing active users (See Figure 3(b)) as well as the number of reports they made (See Figure 3(a)). An average of 226 users were active on the platform each month. Of this number, 123 (54 percent) were new users while 103 (46 percent) were returning users. Correspondingly, an average of 1,036 reports were made each month. Of this number, 830 (80 percent) were submitted by returning users and the remaining 206 (30 percent) by new users. These statistics suggest an active user base on the platform and a steady increase in usage activity throughout its lifespan.

*Geographic analysis.* Figure 4 depicts the geographic distribution of reports and users among various districts in Brussels. Several evident patterns can be observed. The disproportionate distribution of reports and users is most distinctive. Figure 4(a) indicates that there are noticeably fewer reports within the "disadvantaged croissant" at the northwestern sector of the city. In this particular case, the visualization has been exaggerated because the length of streets is not homogenously distributed throughout the city. A fairer depiction of the data based on our standardized indicator shows that the number of reports in certain districts was indeed overrepresented (See Figure 4(b)). A similar pattern can be distinguished in Figure 6(c) where a majority of the users are concentrated in the southeastern sector of the city.

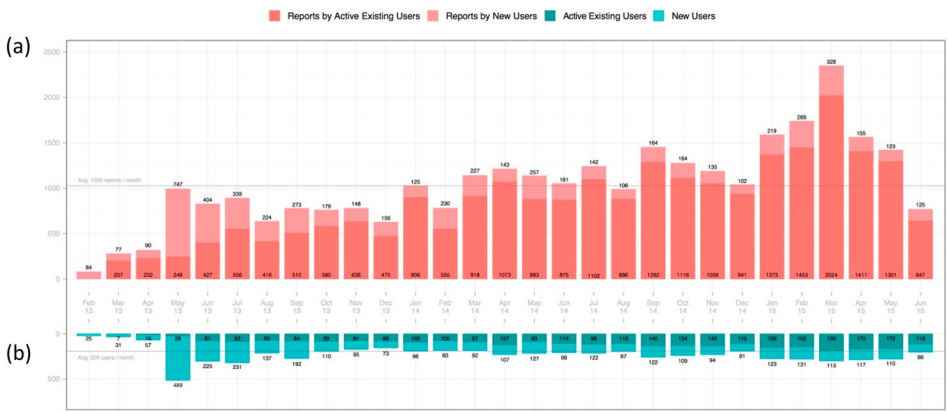

**Figure 3.** Variations in the number of (a) reports made per month by existing active users in comparison to new users, as well as the number of (b) active users in comparison to new users.

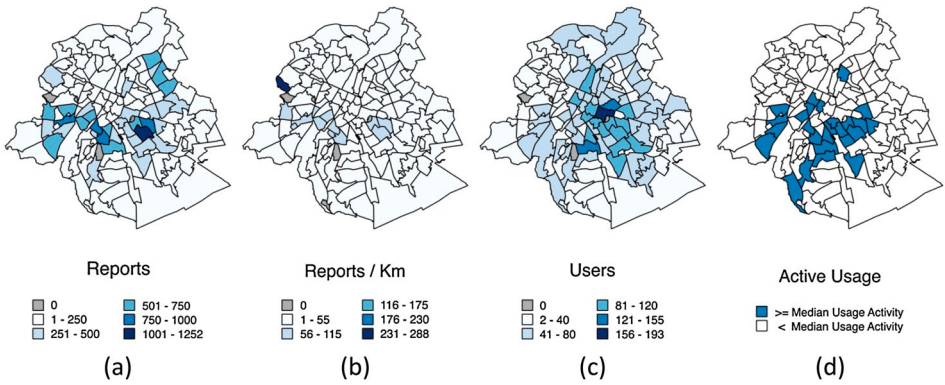

**Figure 4.** The distribution of *FixMyStreet* (a) reports, (b) reports per kilometer, (c) users and (d) districts where users are consistently active.

*FMS* usage per district, summarized in Table 2, indicates that the median number of reports and users per district are 125 and 36, respectively. Additionally, a median of 12 reports is observed for each kilometer of street. Benchmarked against these values, we discovered that only 26 (17 percent) of the districts in Brussels exhibit similar or more active usage patterns. These districts are highlighted in Figure 4(d). The lack of active usage throughout the city is stark and suggests that the platform has not gained traction with a substantial number of inhabitants.

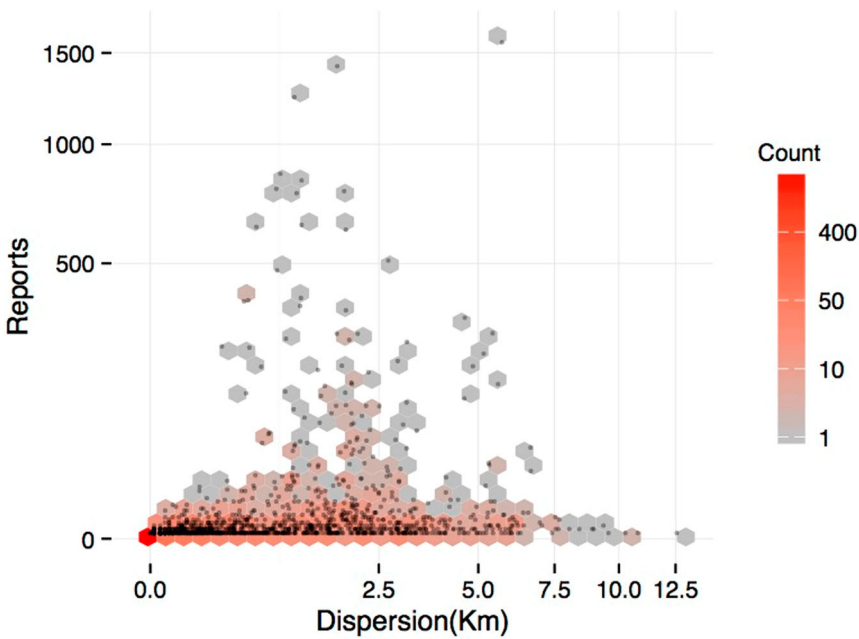

**Figure 5.** The number of *FixMyStreet* reports compared to the average diameter of dispersion for each user on the platform. The scatterplot reveals the presence of several highly active users.

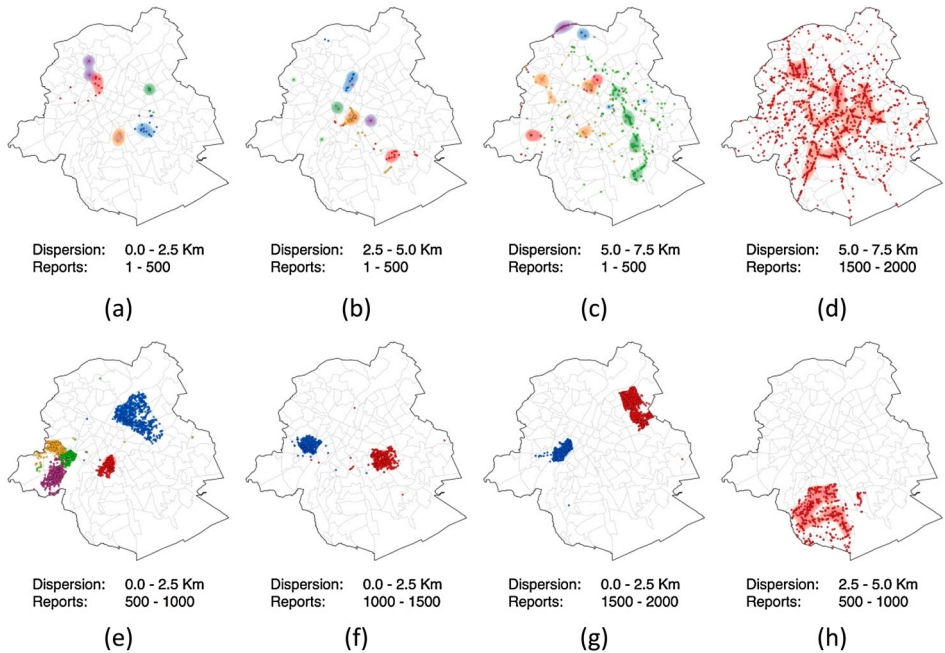

**Figure 6.** Samples of report dispersion patterns that characterize three types of users on *FixMyStreet*. Each map displays up to five sample users. Color is used to differentiate between the sampled users. (a) and (b), first type, is characterized by few reports situated within a relatively small area. (c) and (d), the second type, is highly active and operates over large areas. (e) through (h), the third type, has varying numbers of reports but operate within fixed boundaries.

## Uncovering Different Types of Users

Table 3 contains summary statistics that describe how individual users use the platform. In general, the majority of the users (67 percent) are observed to have reported only one defect. Correspondingly, the average dispersion diameter of 0.471 km suggests that most reported defects occur within a small area.

The scatterplot in Figure 5 corresponds with these statistics though outliers indicate the presence of several highly active users. Three types of users are found upon closer inspection of the data.

In Figure 6, we show samples of report dispersion patterns that characterize each type. Each map depicts up to five sample users. Each dot on the map represents a single report and is color-coded to differentiate between the sampled users. As displayed in Figure 6(a) and 6(b), the first type is observed to submit few reports and operate within a relatively small area.

**Table 2.** Summary of *FixMyStreet* usage per district.

|  | Indicator | $\sigma$ | min | Q1 | med | mean | Q3 | max |
|---|---|---|---|---|---|---|---|---|
| $R$ | Reports | 225.82 | 0 | 68 | 125 | 207 | 250 | 1252 |
| $U$ | Users | 30.43 | 0 | 22 | 36 | 42 | 55 | 193 |
| $R_{u,km}$ | Reports/(User/Km) | 32.12 | 0 | 6 | 12 | 22 | 26 | 288 |

**Table 3.** Summary of individual user activity on *FixMyStreet*.

| Indicator | σ | min | Q1 | med | mean | Q3 | max |
|---|---|---|---|---|---|---|---|
| Reports | 60.96 | 1 | 1 | 1 | 9 | 2 | 1564 |
| Dispersion (Km) | 1.2 | 0 | 0 | 0 | 0.471 | 0.13 | 12.56 |

The dispersion pattern shown in Figure 6(b) differs slightly from Figure 6(a) due to isolated reports outside of certain users' routine reporting area. The second type, as shown in Figure 6(c) and 6(d), is observed to submit many reports over large areas. Some of these users appear to focus on defects that occur along specific transport infrastructures. The third type has varying numbers of reports, yet as illustrated in Figure 6(e)–6(h), their dispersion patterns adhere to certain boundaries. Table 4 indicates that the first type of user is the most common on the platform. Conversely, the second and third types of users are outliers. The relatively high number of type three users within the disadvantaged croissant is noteworthy as it indicates organized activity on the platform. These users could be community volunteers or civil servants.

### Ethnicity and Income

We observed in Figure 1(b) that the average income of citizens living in the northwest and southeast sectors differ substantially. Comparing this to the distribution of ethnic minorities shown in Figure 7(a), we find that the level of income is relatively low in districts where the population density is considerably high (See Figure 7(b)–7(d)). The difference between Weststation and Sint-Pieter aptly characterizes this trend. Whereas only 13 reports were made in Weststation, 988 reports can be observed in Sint-Pieter. What is particularly striking in this comparison is that the number of reports in Sint-Pieter amounts to more than four times the citywide average of 206 per district. This observation verifies earlier work on the status of disadvantaged communities in Brussels (Kesteloot and Meys, 2008), but we identify two anomalies that contradict the overall trend upon closer inspection of the data.

First, we find evidence that suggests citizens of Turkish ethnicity use *FMS* more actively than other disadvantaged ethnic groups. This can be observed in Haachtse Steenweg where 261 reports were submitted. This number is 26 percent higher than the citywide average though the district is primarily occupied by citizens of Turkish ethnicity. Next, the number of reports submitted in Europawijk amounts to 400, almost twice the citywide average though the average income of €2,180 is relatively low. Unlike the former, we believe that this observation can be logically explained by the proximity of Europawijk to the European institutions. In this respect, the majority of the citizens in this district could be expatriates who do not pay taxes and thus are not included in the official statistics. Similarly, commuters who do not live in that district could have been responsible for the reports.

**Table 4.** User types within the "disadvantaged croissant" in comparison to the entire city.

| | Num. of Type 1 Users | Num. of Type 2 Users | Num. of Type 3 Users |
|---|---|---|---|
| Within Disadvantaged Croissant | 647 | 33 | 9 |
| Entire City | 3435 | 65 | 11 |

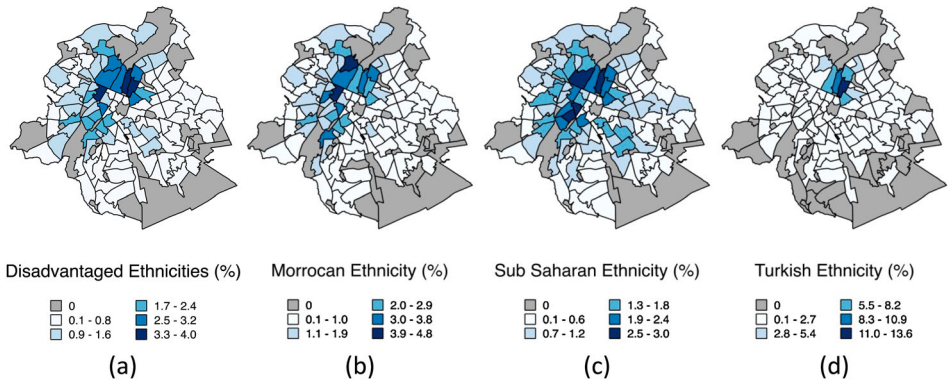

**Figure 7.** The distribution of ethnic minorities based on official statistics. (a) Ethnic minorities aggregated as well as people of (b) North African ethnicity, (c) Sub-Saharan ethnicity, and (d) Turkish ethnicity.

## Languages

Of the 1,134,111 geotagged Tweets in our dataset, 1,063,287 Tweets (94 percent) were classified into 44 languages. French, English, and Dutch are the most commonly used languages in the city and account for 70 percent (744,301 Tweets) of the classified Tweets. These Tweets are homogenously dispersed throughout the city in contrast to those that have been posted in one of the minority languages, which are incidentally grouped into smaller clusters that are dispersed across different districts (see Figure 8 and Table 5).

Focusing solely on the ethnic minorities in Brussels, we discover that Turkish is the most prominent, followed by the Sub-Saharan languages of Somali and Swahili. Arabic on the other hand, is observed to be the least prominent of the three (See Table 5). Although there are more Twitter users who post in Somali and Swahili (Sub-Saharan) than those who tweet in Turkish or Arabic, we observe that Turkish users tweet more on average. Notably, the number of users who tweet in these languages does not correspond proportionately with the population count provided by the official statistics. This is probably a reflection of how the rate of Internet penetration varies among different ethnicities (Quintelier and Vissers, 2008). It is also rational to assume that not all users choose to tweet in their native languages. Another plausible reason for the difference is that alternative patterns that reveal work or leisure venues are captured by Geotagged Tweets since official statistics are based on registered home addresses.

**Table 5.** Summary of Twitter language distribution per district.

|  | Indicators | $\sigma$ | min | Q1 | med | mean | Q3 | max |
|---|---|---|---|---|---|---|---|---|
| $Ar_t$ | Arabic Tweets | 184.51 | 0 | 0 | 2 | 44.99 | 14 | 2032 |
| $SoSw_t$ | Sub-Saharan Tweets | 301.65 | 0 | 46 | 156 | 268.6 | 383 | 1433 |
| $Tr_t$ | Turkish Tweets | 647.12 | 0 | 13 | 60 | 250.4 | 182 | 5620 |
| $Ar_u$ | Arabic Users | 9.92 | 0 | 0 | 1 | 4.193 | 3 | 94 |
| $SoSw_u$ | Sub-Saharan Users | 41.46 | 0 | 18 | 47 | 51.55 | 73 | 211 |
| $Tr_u$ | Turkish Users | 40.58 | 0 | 7 | 20 | 29.45 | 35 | 272 |

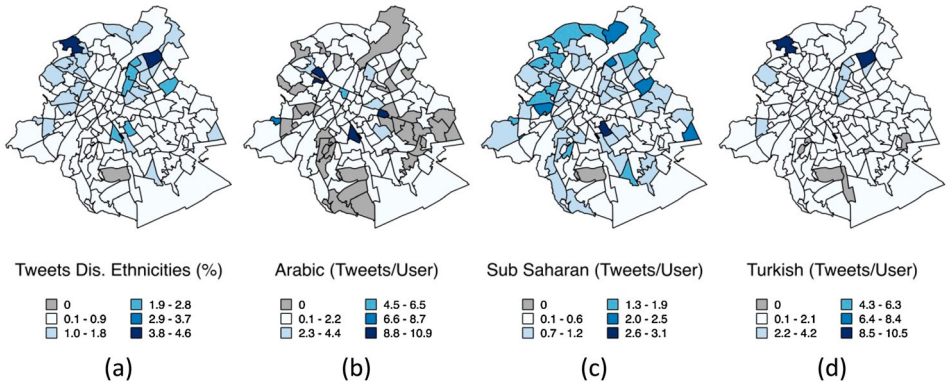

**Figure 8.** The distribution of ethnic minorities approximated with geotagged Twitter data. Presence of (a) aggregated ethnic minorities, (b) North African ethnicity, (c) Sub-Saharan ethnicity, and (d) Turkish ethnicity.

## Hypothesis Testing

Analysis of variance (ANOVA) was applied for hypothesis testing, resulting in five statistically significant outcomes. A detailed breakdown is provided in Table 6.

- **H1** We found a significant difference between average income and the rate of *FMS* participation per district ($p<0.01$). Because the reports were not homogenously distributed across the city, this result suggests that the rate of participation on *FMS* may be influenced by the average income in each district. Nonetheless, this effect does not appear to be systematic as further analysis reveals an insignificant linear correlation (Pearson's $r > 0.05$). Thus, while income indeed influences civic participation on the platform, districts with the highest rate of participation are not necessarily the wealthiest.

**Table 6.** Analysis of variance and the corresponding results of our hypotheses.

| | Hypothesis | Result | F | p | 95% CI |
|---|---|---|---|---|---|
| H1 | There is no significant difference between the rate of FMS participation and average income per district. | R | 8.4 | <0.01** | 11461.552, 9482.498 |
| H2A | There is no significant difference between the rate of FMS participation and the relative number of citizens from disadvantaged ethnicities per district. | R | 5.926 | 0.0162* | 64.772, 46.735 |
| H2B | There is no significant difference between the percentage of Turkish ethnicities and the rate of FMS participation per district. | A | 0.506 | 0.478 | 46.735, 64.772 |
| H2C | There is no significant difference between the percentage of North African ethnicities and the rate of FMS participation per district. | R | 4.349 | 0.0388* | 46.735, 64.772 |
| H2D | There is no significant difference between the percentage of Sub-Saharan ethnicities and the rate of FMS participation per district. | R | 13.59 | <0.01*** | 46.735, 64.772 |
| H2E | There is no significant difference in the rate of FMS participation and the distribution of Tweets written in the languages of disadvantaged ethnicities. | A | 3.821 | 0.0526 | 46.735, 64.772 |
| H2F | There is no significant difference in the rate of FMS participation and the distribution of Turkish Tweets per district. | R | 5.407 | 0.0215* | 46.735, 64.772 |
| H2G | There is no significant difference in the rate of FMS participation and the distribution of Arabic Tweets per district. | A | 0.06 | 0.807 | 46.735, 64.772 |
| H2H | There is no significant difference in the rate of FMS participation and the distribution of Sub-Saharan Tweets per district. | A | 2.304 | 0.131 | 46.735, 64.772 |

- **H2$_A$** We found a significant difference between the relative number of citizens from disadvantaged ethnicities and the rate of *FMS* participation per district ($p<0.05$). Because the reports were not homogenously distributed across the city, this result suggests that the disadvantaged ethnic minorities are underrepresented on the platform.
- **H2$_C$ and H2$_D$** We found a significant difference between the rate of *FMS* participation to the relative number of North African ($p<0.05$) and Sub-Saharan ($p<0.01$) citizens per district respectively. These results indicate that North African and Sub-Saharan citizens are specifically underrepresented on the platform.
- **H2$_F$** We found a significant difference between the rate of *FMS* participation and the relative number of Turkish Tweets per district. This indicates that the rate of Turkish participation on *FMS* does not correspond with their activity on social media.

## Discussion

In summary, our analysis revealed statistically significant differences between the rates of civic participation on *FMS*. Though our findings were specific to Brussels, much of what we found sheds light onto the subject of socio-demographic inequality on web-based civic participation platforms. The results in response to our research questions suggest that web-based civic participation platforms tend to marginalize (RQ1) citizens of certain ethnicities as well as those with (RQ2) lower incomes. These results comport with recent studies. In our context, North Africans and Sub-Saharans were underrepresented on the platform. This outcome complements existing knowledge that describes an overrepresentation of white participants on *FMS* (Gibson et al., 2014). Similarly, this outcome aligns with previous studies that reported a strong bias in participant demographics towards better-educated, high-income social groups (Helsper, 2008; Escher, 2011) that may already be politically engaged (Baykurt, 2012). A possible explanation for this marginalization may be attributed to failures in accommodating different levels of technical aptitude and language proficiency (Baykurt, 2012). The user interface, in particular, might be too complex for all segments of society in Brussels (Offenhuber, 2015). The cost of smart phones and data plans can be an important factor that contributed to the results, considering the ease of use of the smart phone app versus the web application. In addition to these findings, our results suggest that Turkish citizens embrace online social media and actively participate on the *FMS*. We believe that they are more receptive to web-based civic participation as they appear to be more technologically savvy than other ethnic minorities. Turkey has a high Twitter penetration rate (third in the world) (Statista, 2016). A significant part of the Turkish people in Belgium have double nationality, and they are called "summer Turks" by the permanent residents of Turkey. They frequently travel to their hometowns and get heavily influenced by trends in Turkey.

These findings are not without limitations. From a counter-perspective, the unequal distribution of reports in our specific context may be due to late adoption of the platform by citizens in ethnically diverse and deprived districts. As discussed in the former section, the communication channels used for the promotion of *FMS* were not necessarily accessible by people of all socioeconomic levels and national origins. Thus lower rates of participation in these districts may change over time. Nonetheless, this delayed adoption raises questions about the intention of the government to include citizens living in deprived districts on the platform. The long-term effects of equal participation rates may lead to an increase in the quality of the physical environment. In contrast, unequal participation in ethnically diverse deprived districts may result in the unequal treatment of defects in the physical environment and potentially amplify the impression of sociodemographic inequalities in cities. While somewhat speculative, our analysis suggests the presence of community volunteers or civil servants who operate within deprived districts. If this insight proves to be true, it indicates positive government intervention towards promoting the platform in these districts. However, this intervention should be clarified in the data and depicted accordingly on the user interface, as it may give a misleading impression of homogenous participation rates across the city. The results should also be interpreted cautiously due to the aggregation of *FMS* reports into districts that may not necessarily reflect how citizens perceive their neighborhoods. This was necessary to compare the *FMS* data to the official statistics (BISA, 2016) that were collected at the district level. Although these districts were designated by social scientists to represent a specific demographic cluster, the aggregation of *FMS* points in these districts fails to capture the continuity and finesse of everyday life such as the streets heavily inhabited by different demographics. Furthermore, the automatic language categorization used in this study may not be able to match the accuracy of human classification due to the informal style of writing or when multiple languages are mixed in a single Tweet.

This work presents implications for the design and implementation of similar platforms in the future. We believe that participation on web-based civic participation platforms may indeed require skills that are beyond the reach of disadvantaged groups. Until computers and Internet access become more affordable, web-based civic participation platforms like *FMS* may be best suited as a supplement for more traditional channels to engage the public. A possible way to address this challenge would be to develop special programs that support disadvantaged communities to improve their skills and promote the benefits of web-based civic participation. As a recommendation for future implementations of similar platforms, we stress the importance of understanding the sociodemographic composition of the local context. Rather than following a "one size fits all" approach, the overall design, representation styles, interface, access modes, and language of the system should be carefully tuned to be as inclusive as possible. Another important point to note is that FMS is very far from government searches designed to find potholes. It should rather be thought as a complementary effort to the government searches for potholes.

As a future direction, solutions for the automatic recognition of problems from the submitted photographs has yet to be explored. Santos et al. (2013) have already developed a basis for a system capable of identifying the most common problems spotted in images. This would significantly improve the usability of *FMS*, by effectively and efficiently processing and forwarding instances correctly. Besides the automatic processing of the photos, a possible improvement direction for *FMS* is to employ second generation open data initiatives such as *StreetBump* (Lee et al., 2015), which are capable of collecting data about the road conditions as citizens drive (or bike). Using this method, it is possible for cities to aggregate data about real-time road problems without the need for the direct interaction with the users. Another interesting potential to be explored in the future is the possible link between automobile ownership and *FMS* reports. It is logical to expect that potholes on the roads might not be so important for a person without a car. This research can as well be extended to the "Cambio" shared car use points which are quite active in the Brussels Capital Region in relation to the income levels and demographics of the users.

## Conclusion

We have analyzed the historical usage logs from *FMS*, official city statistics, and geotagged social media data from Twitter to investigate socio-demographic inequality on web-based civic participation platforms. Our analysis, based on a case study situated in Brussels led to several statistically significant findings that suggest that web-based civic participation platforms tend to marginalize citizens of certain ethnicities as well as those with lower incomes. In this respect, our findings provide timely evidence to inform the design of more inclusive crowdsourced, civic participation platforms in the future.

## Disclosure Statement

No potential conflict of interest was reported by the authors.

## Notes on Contributors

*Burak Pak* is a postdoctoral researcher and design studio teacher in the Faculty of Architecture, KU Leuven. His research focuses on alternative uses of digital spatial media for design empowerment which follows two complementary tracks of enabling bottom-up participation (in) and (through) research and design.

*Alvin Chua* holds a PhD from the Department of Architecture, KU Leuven. He is currently employed at the Urban Redevelopment Authority of Singapore (URA) as a data scientist. His research interests lie in urban planning, urban informatics, geovisualization, data visualization and visual analytics.

*Andrew Vande Moere* is an associate professor in the Department of Architecture, KU Leuven. He is the vice head of the Research[x]Design (RxD) research group. His research interests include emerging design technologies, human-computer interaction, architectural and urban computing, media architecture, information visualization and data-driven urban design.

## ORCID

*Burak Pak*  http://orcid.org/0000-0002-2901-1032
*AlvinChua*  http://orcid.org/0000-0002-0813-3743